\documentclass[useAMS,usenatbib,usegraphicx]{mn2e}

\newcommand{\expect}[1]{\langle #1 \rangle}     % expectation value
\newcommand{\lexpect}[1]{\left\langle #1 \right\rangle}  % expectation value
\newcommand{\Mpc}{$h^{-1}$\,Mpc}
\newcommand{\iMpc}{$h$\,Mpc$^{-1}$}

\newcommand{\bk}{{\bmath k}}
\newcommand{\br}{{\bmath r}}

\newcommand{\art}{{\sc art}}
\newcommand{\grafic}{{\sc grafic}}
\newcommand{\nbody}{$N$-body}
\newcommand{\matF}{\mathbfss{F}}

\newcommand{\matU}{\mathbfss{U}}
\newcommand{\matW}{\mathbfss{W}}
\newcommand{\matD}{\mathbfss{D}}
\newcommand{\matL}{\mathbf\Lambda}
\newcommand{\matP}{\mathbfss{P}}

\title[Information in the non-linear power spectrum]{Information content of the
  non-linear power spectrum: the effect of beat-coupling to large scales}
  \author[C. D. Rimes and A. J. S. Hamilton] {Christopher
  D. Rimes$^1$\thanks{E-mail: rimes@colorado.edu, Andrew.Hamilton@colorado.edu}
  and Andrew J. S. Hamilton$^{1, 2}$$^\star$\\ $^1$JILA, University of
  Colorado, 440 UCB, Boulder, CO 80309-0440, U.S.A.\\ $^2$Department of
  Astrophysical and Planetary Sciences, University of Colorado, 391 UCB,
  Boulder, CO 80309-0391, U.S.A.}

\begin{document}

%\date{For submission to Monthly Notices of the Royal Astronomical Society}
%\date{Submitted to Monthly Notices of the Royal Astronomical Society}
\date{Accepted ... Received ...; in original form ...}

\pagerange{\pageref{firstpage}--\pageref{lastpage}} \pubyear{2005}

\maketitle

\label{firstpage}

\begin{abstract}
We measure the covariance of the non-linear matter power spectrum from \nbody\
simulations using two methods.  In the first case, the covariance of power is
estimated from the scatter over many random realizations of the density field.
In the second, we use a novel technique to measure the covariance matrix from
each simulation individually by re-weighting the density field with a carefully
chosen set of functions.  The two methods agree at linear scales, but
unexpectedly they disagree substantially at increasingly non-linear scales.
Moreover, the covariance of non-linear power measured using the re-weightings
method changes with box size.  The numerical results are consistent with an
explanation given in a companion paper, which argues that the cause of the
discrepancy is beat-coupling, in which products of Fourier modes separated by a
small wavevector couple by gravitational growth to the large-scale beat mode
between them.  We calculate the information content of the non-linear power
spectrum (about the amplitude of the initial, linear power spectrum) using
both methods and confirm the result of a previous paper, that at translinear
scales the power spectrum contains little information over and above that in
the linear power spectrum, but that there is a marked increase in information
at non-linear scales.  We suggest that, in real galaxy surveys, the covariance
of power at non-linear scales is likely to be dominated by beat-coupling to the
largest scales of the survey and that, as a result, only part of the
information potentially available at non-linear scales is actually measurable
from real galaxy surveys.
\end{abstract}


%%%%%%%%%%%%%%%%%%%%%%%%%%%%%%%%%%%%%%%%%%%%%%%%%%%%%%%%%%%%%%%%%%%%%%%%%%%%%%%
%%%%%%%%%%%%%%%%%%%%%%%%%%%%%%%%%%%%%%%%%%%%%%%%%%%%%%%%%%%%%%%%%%%%%%%%%%%%%%%

\section{Introduction}
\label{sec: introduction}

Recent progress in cosmological parameter estimation has been characterized by
rapid convergence of constraints from a wealth of different types of
observations -- galaxy clustering, the Lyman-$\alpha$ forest power spectrum,
galaxy cluster abundances, high-redshift type Ia supernovae, weak gravitational
lensing, big-bang nucleosynthesis and many others -- toward a single,
well-determined `concordance' cosmological model.  Of particular importance has
been the combination of high precision maps of anisotropies in the cosmic
microwave background (CMB) with measurements of galaxy clustering from large
redshift surveys, which yield complementary constraints on key cosmological
parameters (e.g.\ \citealt{Efstathiou02}; \citealt{Spergel03};
\citealt{Tegmark04a}; \citealt{Seljak05}; \citealt{Sanchez05}).

Galaxy clustering analyses (e.g.\ \citealt{Percival01}; \citealt{Tegmark04a};
\citealt{Cole05}; \citealt{Eisenstein05}) are generally restricted to scales
$\ga 20$~\Mpc, where density fluctuations are still linear, galaxy-matter bias
appears to be independent of scale (although it does depend on luminosity and
on galaxy type) and the matter power spectrum directly traces the spectrum of
density fluctuations at recombination.  At smaller scales, where much of the
observational data in galaxy surveys lie, the extent to which the linear power
spectrum can be recovered from the non-linear power spectrum remains unknown.
Non-linear evolution changes the shape of the power spectrum in a non-trivial
way, and introduces broad correlations between measurements of power at
different wavenumbers (\citealt{MW99}; \citealt*{SZH99}; \citealt{CH01}).  The
early success of analytic formalisms at producing invertible one-to-one
mappings between linear and non-linear spectra (\citealt{HKLM91};
\citealt[1996\nocite{PD96}]{PD94}) suggested that information in the linear
power spectrum may be preserved into the non-linear regime, but other work
(\citealt*{MWP99}; \citealt{SE05}) has shown that non-linear evolution erases
features in the power spectrum, perhaps leading to an irreversible loss of
information.

In an earlier letter \citep[hereafter Paper~I]{Paper1}, we reported
measurements of the amount of information in the non-linear power spectrum
about the amplitude of the linear power spectrum for the currently favoured
(concordance) cosmological model, from a large ensemble of \nbody\ simulations.
We have since discovered a small error in our calculations.  In Fig.~\ref{fig:
info} of the present paper we present a revised version of the results from
Paper~I.  Our conclusions remain unchanged: namely, that there exists little
independent information in the translinear regime ($k \simeq 0.2$--0.8~\iMpc\
at the present day) over and above that in the linear power spectrum, but that
in the fully non-linear regime there appears to be a significant amount of
information beyond that measurable from the linear power spectrum.

Measuring the information content of the non-linear power spectrum involves
measuring the covariance matrix of non-linear power, which, if determined as in
Paper~I from the scatter over an ensemble, requires performing a large number
of \nbody\ simulations.  This is costly in terms of computing time, especially
if the simulations are of high quality.  In an attempt to reduce the
computational overhead and to streamline the measurement of information, we
devised a new technique, described in detail in a companion paper
\citep*[hereafter HRS]{HRS05}, for estimating the covariance of power from
individual simulations, by re-weighting the density field using a set of
carefully chosen windows.

In the present paper, we use the re-weighting technique to measure the amount
of information in the non-linear power spectrum, for the same set of
simulations used in Paper~I.  We compare the results to those obtained with
the ensemble method and present a number of tests of the method.

Unexpectedly, we find that, far from agreeing, the covariance of power measured
by the re-weightings method substantially exceeds that measured by the ensemble
method.  When we first encountered this discrepancy, it seemed to us that it
must be caused by a `bug' in our code, and we performed numerous numerical
tests to track it down.  Belatedly, we realized that the discrepancy was caused
not by a bug but by a real physical process, which we term `beat-coupling'.
The physical origin of beat-coupling is described by HRS, who illustrate its
effects with examples using perturbation theory and the hierarchical model.

This paper is organized as follows.  In Section~\ref{sec: info} we set out our
definition of information, and discuss the decorrelation choices that must be
made to allocate information to prescribed wavebands.  This section also
provides a more detailed exposition of the techniques employed in Paper~I.
Section~\ref{sec: simulations} describes the numerical simulations used in both
this paper and the previous one.  Section~\ref{sec: covariance} compares
measurements of the covariance of power using both ensemble and re-weightings
methods, and describes several tests of the results.  In Section~\ref{sec:
results} we compare the information content of the non-linear power spectrum
measured using the two different methods.  Our conclusions are summarized in
Section~\ref{sec: summary}.

%%%%%%%%%%%%%%%%%%%%%%%%%%%%%%%%%%%%%%%%%%%%%%%%%%%%%%%%%%%%%%%%%%%%%%%%%%%%%%%
%%%%%%%%%%%%%%%%%%%%%%%%%%%%%%%%%%%%%%%%%%%%%%%%%%%%%%%%%%%%%%%%%%%%%%%%%%%%%%%

\section{Information}
\label{sec: info}

\subsection{Fisher information}
\label{ssec: fisher}

The Fisher information matrix is defined \citep*[e.g.][]{TTH97} as
\begin{equation}
  F^{\alpha\beta} \equiv - \lexpect{\frac{\partial^2{\ln\mathcal{L}}}
                                   {\partial{p_\alpha}\partial{p_\beta}}},
\end{equation}
where $\mathcal{L}(p_\alpha | \mbox{data})$ is the likelihood function -- the
multivariate probability distribution of the model parameters $p_\alpha$ given
the available data and a set of model assumptions (the Bayesian prior).  Fisher
information is additive over independent measurements, clearly a desirable
property for information to possess.  Its importance in parameter estimation is
encapsulated in the Cram\'er-Rao inequality, which limits the maximum precision
with which a single parameter $p_\alpha$ can be measured to
\begin{equation}
  \expect{\Delta{\hat{p}_\alpha}^2} \ge 1/F_{\alpha\alpha},
\end{equation}
if the estimator $\hat{p}_\alpha$ is unbiased and if this is the only parameter
being estimated from the data.  Here and throughout this paper we use hats to
distinguish an \emph{estimate} of a quantity from its true value.  If estimates
of the parameters are Gaussian distributed about their expectation values -- a
good approximation in the limit of a large amount of data, thanks to the
central limit theorem -- then their covariance matrix is well-approximated by
the inverse of the Fisher matrix:
\begin{equation}
  \expect{\Delta\hat{p}_\alpha\Delta\hat{p}_\beta} \simeq
  (F^{-1})_{\alpha\beta}.
\end{equation}

\subsection{Power spectrum}
\label{ssec: power}

We consider a statistically homogeneous and isotropic density field
$\rho(\br)$.  The power spectrum $P(k)$ of density fluctuations of such a field
is defined by
\begin{equation}
  \expect{\delta_\bk\delta_{\bk^\prime}} = (2\upi)^3\delta_{3D}(\bk+\bk^\prime)P(k),
\end{equation}
where $\delta_\bk$ is the Fourier transform of the overdensity $\delta(\br)
\equiv \rho(\br)/\bar{\rho}-1$ and $\delta_{3D}(\bk)$ is a 3-dimensional Dirac
delta function.  Statistical isotropy requires that the power spectrum be a
function only of the magnitude $k \equiv |\bk|$ of the wavevector $\bk$.

For Gaussian fluctuations, each $\delta_\bk$ has real and imaginary components
that are independently Gaussianly distributed with variance $P(k)/2$.  Usually,
power is estimated by averaging over shells in Fourier space.  For Gaussian
fluctuations, the expected covariance matrix of shell-averaged estimates of
power is diagonal, with variance
\begin{equation}
  \expect{\Delta{\hat{P}(k)}^2} = 2P(k)^2/N_k,     \label{eq: variance}
\end{equation}
where $N_k$ is the number of modes in the shell around $k$ (a finite number in
the case of a realistic galaxy survey or a periodic \nbody\ simulation).  Here
$\delta_\bk$ and its complex conjugate $\delta_{-\bk}$ are counted as
contributing two distinct modes, the real and imaginary parts of $\delta_\bk$.

\subsection{Information in the power spectrum}
\label{ssec: power info}

Here, as in Paper~I, we measure the Fisher information $I$ in a single
parameter: the log of the amplitude $A$ of the initial (post-recombination)
matter power spectrum, that is,
\begin{equation}
\label{eq: I}
  I \equiv - \lexpect{ \frac{\partial^2\ln\mathcal{L}}{\partial\ln{A}^2} }.
\end{equation}
For Gaussian density fluctuations, the power spectrum completely specifies the
statistical properties of the density field, so that the only explicit
dependence of the likelihood ${\cal L}$ is on the power spectrum.  For
non-Gaussian fluctuations, the likelihood function may also depend explicitly
on other parameters.  However, as in Paper~I, we consider only the information
contained in the power spectrum $P(k)$, in which case the information $I$
defined by equation~(\ref{eq: I}) can be expanded as
\begin{equation}
\label{eq: IPk}
  I = - \left\langle \sum_{k,k^\prime} \frac{\partial\ln P(k)}{\partial\ln A}
  \frac{\partial^2\ln\mathcal{L}}{\partial\ln P(k) \, \partial\ln
  P({k^\prime})} \frac{\partial\ln P({k^\prime})}{\partial\ln A} \right\rangle.
\end{equation}
The middle term on the right-hand side of equation~(\ref{eq: IPk}) is the
Hessian of the vector $\ln P(k)$ of log non-linear powers, the expectation
value of which is the Fisher matrix.  The power spectrum $P(k)$ is averaged
over spherical shells in $k$-space; a typical shell in the non-linear regime
contains several thousand to hundreds of thousands of distinct Fourier modes,
so it is reasonable to invoke the central limit theorem to assert that
estimates of power will be Gaussianly distributed about their expectation
values.  This assertion holds even if the density field is itself non-Gaussian.
We test this is explicitly in Section~\ref{sssec: gaussian}.  In the Gaussian
approximation, the Hessian in equation~(\ref{eq: IPk}) can be approximated by
the inverse of the covariance matrix of estimates of log-power.

The remaining terms in equation~(\ref{eq: IPk}) are two partial derivatives
which describe the sensitivity of the non-linear power to changes in the
amplitude $A$ of the initial linear power.  In the linear regime these
derivatives are identically unity, since $P_{\rm L}(k) \propto A$; at
non-linear scales they are equal to the growth rate of the non-linear power
spectrum relative to the linear, which can be conveniently measured from
simulations.

The information $I$ has a particularly simple interpretation for Gaussian
fluctuations.  Following equation~(\ref{eq: variance}), it is equal to half the
total number $N$ of Gaussian modes:
\begin{equation}
  I = N/2.
\end{equation}

As was found in Paper~I and is confirmed in Section~\ref{sec: results} of the
present paper, the information in the non-linear power spectrum $P(k)$ is
significantly less than the information in the linear power spectrum at the
same wavenumber.  This decrease in information could result from a transfer of
information from larger to smaller scales, a diversion of information into
other quantities (such as the bispectrum), or an irreversible loss of
information.  In Paper~I we argued that complete loss of information during
translinear evolution is inconsistent with our finding that the total amount
of information on non-linear scales is increasing with time.  The remaining two
scenarios could, in principle, be distinguished by measuring the information
the bispectrum and higher order statistics but this becomes progressively more
difficult as the order increases.

\subsection{Decorrelated band powers}
\label{ssec: decorrelating}

The quantity $I$ defined by equation~(\ref{eq: IPk}) is the total amount of
information contained in the non-linear power spectrum about the parameter $\ln
A$.  Some of this information is degenerate between measurements of power at
different wavenumbers as a result of the broad correlations introduced by
non-linear evolution.

\citet{HT2000} showed how to decorrelate a power spectrum by defining a set of
windowed band-power estimates.  Decorrelation is the process of assigning shared information
uniquely to a given wavenumber.  Here, we extend their method to the case where
we want to decorrelate, not the power spectrum itself, but estimates of some
parameter -- in this case $\ln A$ -- made from the power spectrum.

Following \citet{HT2000}, we define our windowed band-power estimates
$\hat{B}_k$ by:
\begin{equation}
  \frac{\hat{B}_k}{P_k} = \sum_{k^\prime} W_{kk^\prime}
          \frac{\hat{P}_{k^\prime}}{P_{k^\prime}},
\label{eq: band powers}
\end{equation}
where we use the index notation $P_k=P(k)$ to emphasize the fact that the
shell-averaged power spectrum $P(k)$ can be viewed as a discrete vector in
Fourier space.  The band-power windows $W_{kk^\prime}$ in equation~(\ref{eq:
band powers}) are elements of a decorrelation matrix, each column of the matrix
being a discrete window for one band-power $B_k$.  There are many (actually an
infinite number) of schemes for decorrelating the power spectrum, corresponding
to different ways of sharing out degenerate information between wave bands.
The reader is directed to \citet{HT2000} for a discussion of the relative
merits of selected decorrelation schemes.

Both sides of equation~(\ref{eq: band powers}) are scaled by $P_k$, which is a
fiducial power spectrum.  This scaling ensures that a given band power $B_k$ is
not dominated by leakage from wavenumbers $k^\prime$ where the window is small
but $P_{k^\prime}$ is large.  The choice $P_k=\expect{\hat{P}_k}$ guarantees
that the expectation value of the windowed band power estimates at each
wavenumber is equal to the original power spectrum, provided that the windows
are suitably normalized:
\begin{equation}
  \sum_{k^\prime} W_{kk^\prime} = 1.
\end{equation}

In order that the final estimates of $\ln A$ are uncorrelated, the band-power
windows must satisfy
\begin{equation}
  \matW^\top\matL\matW = \matD\matF\matD,
\end{equation}
where $\matF$ is the Fisher matrix of the scaled power spectrum and $\matD$ is
a diagonal matrix with diagonal elements
\begin{equation}
  D_{ii} = \frac{\partial\ln P_{k_i}}{\partial\ln A}.    \label{eq: matrix D}
\end{equation}
The matrix $\matD\matF\matD$ can be interpreted as the Fisher matrix of
estimates of $\ln A$ from the power in different wavebands and it is these
estimates (rather than the estimates of power themselves) that we want to
decorrelate.  We experimented with various decorrelation matrices, eventually
opting for the upper triangular matrix $\matU$ obtained from a generalized form
of the Cholesky decomposition:
\begin{equation}
  \matU^\top\matL\matU = \matD\matF\matD, \label{eq: cholesky}
\end{equation}
where $\matL$ is a diagonal matrix -- the Fisher matrix of the decorrelated
estimates of $\ln A$.  Note that this is \emph{not} the same as decorrelating
the power spectrum using the Cholesky decomposition of $\matF$ and then writing
the Fisher matrix of the decorrelated estimates as $\matD\matL\matD$ -- as we
did in Paper~I -- because $\matD$ does not commute with $\matU$.  The two
approaches are only approximately equivalent in the case where the band-power
windows are narrow or the elements of $\matD$, given by equation \ref{eq:
matrix D}, are a slowly varying function of $k$.  The former is a particularly
poor approximation, because of the broad correlations present in the non-linear
power spectrum.  In Section~\ref{sec: results}, we present a corrected version
of the relevant figure (fig.~3) from Paper~I.  Our conclusions are not altered
significantly.

In the central-limit-theorem approximation that estimates of power are
Gaussianly distributed, the Fisher matrix $\matF$ is approximately equal to the
scaled inverse covariance of power:
\begin{equation}
  \matF \simeq \matP\expect{\Delta\hat\matP\Delta\hat\matP^\top}^{-1}\matP^\top.
\end{equation}
Here, $\matP$ is a diagonal matrix whose non-zero elements are equal to the
fiducial power spectrum $P_k$.  Mathematically, upper Cholesky decorrelation is
equivalent to taking a matrix composed of all the elements of the covariance
matrix up to some wavenumber $k_{\rm max}$, inverting this matrix, and summing
all the elements of the resulting Fisher matrix to arrive at a measure of the
accumulated information $I(\leq k_{\max})$ up to that wavenumber.

Upper Cholesky decorrelation yields band-power windows that are highly
asymmetric, with the band power at each wavenumber $k$ containing contributions
only from power on larger scales.  A problem with the other schemes that we
tried (including the square root of the scaled Fisher matrix, recommended by
\citealt{HT2000}), is that there is an appreciable covariance between large,
linear scales and small, non-linear scales.  Applying anything other than upper
Cholesky decorrelation assigns some of this covariance to large scales, causing
the information at linear scales to depart from the expected Gaussian
information, equation~(\ref{eq: variance}).

While there is a certain arbitrariness about choosing upper Cholesky
decorrelation over other possibilities, the resulting cumulative information
$I(\leq k_{\max})$ has the virtue of a simple interpretation: it is the
information in the power spectrum $P(k)$ at wavenumbers $k \leq k_{\max}$,
uncontaminated by information in power at smaller scales.

Because the Fisher matrix of the uncorrelated band powers $B(k)=B_k$ is by
definition diagonal, equation~(\ref{eq: IPk}) reduces to a sum over a single
wavenumber and the cumulative information is:
\begin{equation}
  I(\le k) = - \left\langle \sum_{k=0}^{k_{\rm max}} \frac{\partial\ln B(k)}{\partial\ln A}
      \frac{\partial^2\ln\mathcal{L}}{\partial\ln B(k)^2} \frac{\partial\ln
      B(k)}{\partial\ln A} \right\rangle.
	  \label{eq: IBk}
\end{equation}

%%%%%%%%%%%%
% FIGURE 1 %
%%%%%%%%%%%%

\begin{figure}
\centerline{\includegraphics[width=8cm]{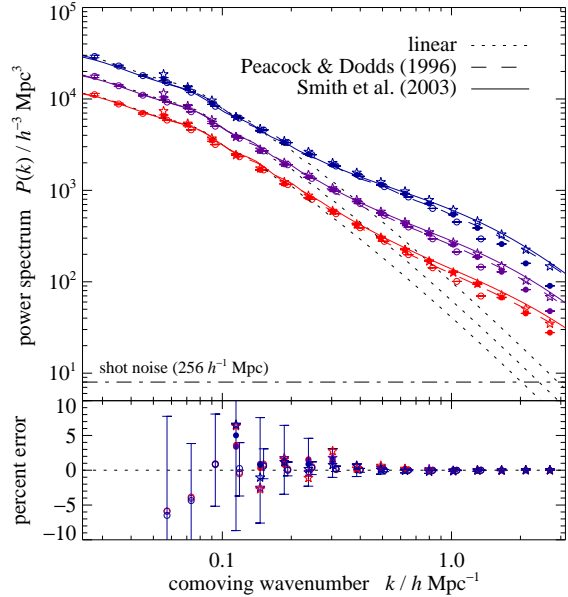}}
\caption{Evolution of the non-linear power spectrum.  Top panel: mean power
  spectrum from the 256~\Mpc\ PM simulations (open points, with error bars
  derived from the scatter between individual simulations); the 128~\Mpc\ PM
  simulations (filled points with error bars); and the 128~\Mpc\ \art\
  simulations (stars).  Power spectra are shown for three epochs (bottom to
  top: $a=0.5$, 0.67 and 1).  The linear power spectrum is shown by the dotted
  curves in each panel.  The solid and dashed curves are, respectively, from
  the fitting formulae of \citet{Smith03} and \citet{PD96}. The dot-dashed line
  marks the level of the shot noise in the 256~\Mpc\ simulations; the shot
  noise in the 128~\Mpc\ simulations is a factor of $8$ lower.  Bottom panel:
  deviation between the mean power spectra of weighted densities and the power
  spectrum of the unweighted density.  The points are medians from 100 of the
  PM simulations of each box size and the 25 \art\ simulations at the same 3
  epochs.  Error bars mark the upper and lower quartiles of the distribution.
  \label{fig: power}}
\end{figure}

%%%%%%%%%%%%%%%%%%%%%%%%%%%%%%%%%%%%%%%%%%%%%%%%%%%%%%%%%%%%%%%%%%%%%%%%%%%%%%%
%%%%%%%%%%%%%%%%%%%%%%%%%%%%%%%%%%%%%%%%%%%%%%%%%%%%%%%%%%%%%%%%%%%%%%%%%%%%%%%

\section{Simulations}
\label{sec: simulations}

The simulations used in this paper are the same as those used in Paper~I.  The
main ensemble comprises 600 gravitational \nbody\ simulations of the
concordance cosmological model: 400 simulations with a box size of 256~\Mpc\
and a further 200 with a box size of 128~\Mpc.  These simulations were evolved
using a particle-mesh (PM) code with $128^3$ dark matter particles and a
$256^3$ force mesh.

An additional 25 simulations, also with a box size of 128~\Mpc, were run using
a parallel version of the adaptive mesh refinement (AMR) code \art\
\citep*{KKK97}.  Alone, this is an insufficient number to give precise
statistics, but together with the much larger ensemble of PM simulations, the
\art\ simulations serve as a useful check of our results on small scales, where
the AMR technique is more accurate.  For the \art\ simulations, we used $128^3$
particles and a $128^3$ root mesh with, at most, three levels of refinement,
giving a maximum spatial resolution equivalent to that of a $1024^3$ mesh in
dense regions.  Gaussian initial conditions for the \art\ simulations were set
up using the publicly available \grafic\ package.

The cosmological parameters adopted are those of \citet[second-last column of
their table 4]{Tegmark04a}:
\begin{equation}
  (\Omega_{\rm M},\Omega_{\Lambda},f_{\rm b},h,\sigma_8) = (0.29,0.71,0.16,0.71,0.97)
\end{equation}
This choice of parameters gives the best fit to the combination of the power
spectrum of fluctuations in the cosmic microwave background as measured by the
Wilkinson Microwave Anisotropy Probe (WMAP) and galaxy clustering as measured
by the Sloan Digital Sky Survey (SDSS), under the important assumptions (among
others) of a spatially flat universe ($\Omega_k=0$) with a cosmological
constant ($w=-1$) and a scale-invariant primordial power spectrum ($n_{\rm
s}=1$).  Each simulation was seeded with a different, randomly chosen
realization of a Gaussian random field with a power spectrum corresponding to
the above cosmological model.  The matter transfer function was calculated from
the fitting formula of \citet{EH98} for universes with a significant baryon
content.

The non-linear power spectra of the above simulations are plotted in
Fig.~\ref{fig: power} for three epochs: $a=0.5$, 0.67 and 1.0, with error bars
showing the scatter between individual realizations.  Power spectra were
calculated by Fourier transforming the weighted density field on a single
$256^3$ grid\footnote{`Chaining' \citep{Jenkins98}, a clever way to extend
measurements of the power spectrum to smaller scales by superposing the eight
octants of a periodic cube on to a single octant, unfortunately cannot be
applied to the measurement of covariance, because it involves a reduction in
the number of Fourier modes.  For Gaussian fluctuations, each mode is
independent, so each folding reduces the number of modes by a constant factor
8.  At non-linear scales, however, adjacent modes are highly correlated, so
subsampling them by a factor of 8 does not reduce the effective number of modes
correspondingly.}.  The resulting power spectra were corrected for smoothing by
dividing by the square of the Fourier transform of the mass assignment window,
prior to subtracting the shot noise contribution \citep{Smith03}.  As is to be
expected, the power in the PM simulations falls significantly below that
measured in the higher resolution \art\ simulations at small scales, as a
result of mesh effects in the PM simulations.  The power spectrum from the
\art\ simulations, on the other hand, agrees well with the fits to previous
\nbody\ simulations by \citet{PD96} and \citet{Smith03} at the scales of
interest.  We restrict our analyses to scales for which corrections to power
from particle-cell assignment are small, and the shot noise sub-dominant, so
uncertainties in both of these corrections should not affect our results.

In what follows, where the results from the 128~\Mpc\ PM and \art\ simulations
are consistent, we combine them into a single data set for clarity.

%%%%%%%%%%%%%%%%%%%%%%%%%%%%%%%%%%%%%%%%%%%%%%%%%%%%%%%%%%%%%%%%%%%%%%%%%%%%%%%
%%%%%%%%%%%%%%%%%%%%%%%%%%%%%%%%%%%%%%%%%%%%%%%%%%%%%%%%%%%%%%%%%%%%%%%%%%%%%%%

\section{Constructing the covariance matrix}
\label{sec: covariance}

Estimating the amount of information in a set of measurements requires
knowledge of their covariance matrix.  The most direct way to measure the
covariance of estimates of the power spectrum is to run an ensemble of \nbody\
simulations, each having a different, random realization of the initial density
field (e.g.\ \citealt{MW99}; \citealt{SZH99}; Paper~I) -- a computationally
expensive endeavour because many hundreds of realizations are required to yield
an accurate estimate of the covariance matrix (\citealt{MW99}; Paper~I).

Alternative approaches to measuring covariances include `jackknife', in which
ensembles are formed by removing selected data from the original sample, and
`bootstrap', in which ensembles are formed by resampling with replacement
\citep{Kunsch89}.  These methods work provided that the data being sampled
comprise independent random variables drawn from the same distribution.  For
correlated data, \citet{Kunsch89} suggested an extension to the bootstrap
approach, in which the data are first split into blocks whose length is larger
than the characteristic length of the correlations, and these blocks are then
re-sampled to generate the bootstrap sample.  In early experiments, we tried a
form of `block bootstrap' in which we filled each octant of a simulation cube
with a block of data selected randomly from the cube.  Unfortunately, the
method did not work well.  The sharp edges of the octants introduced spurious
small-scale power, and the covariance of small-scale power differed
substantially from that measured by the ensemble method.  The mathematical
relation between the covariance of power obtained by the block bootstrap method
and the true covariance of power is sufficiently obscure that it was difficult
to assess the possible causes of the discrepancy.

In HRS, we argue that all variations on jackknife and bootstrap are really just
different ways of re-weighting the data to yield a new estimate of the quantity
of interest, and that the best way to avoid unpleasant side-effects on
spatial data is to re-weight with a smooth function of position.  Further, by
re-weighting the simulation with a function that contains only large-scale
Fourier modes, unpleasant numerical artefacts in the power spectrum are
confined to the largest scales, making them much easier to deal with.

%%%%%%%%%%%%
% FIGURE 2 %
%%%%%%%%%%%%

\begin{figure*}
\centerline{\includegraphics[width=16cm]{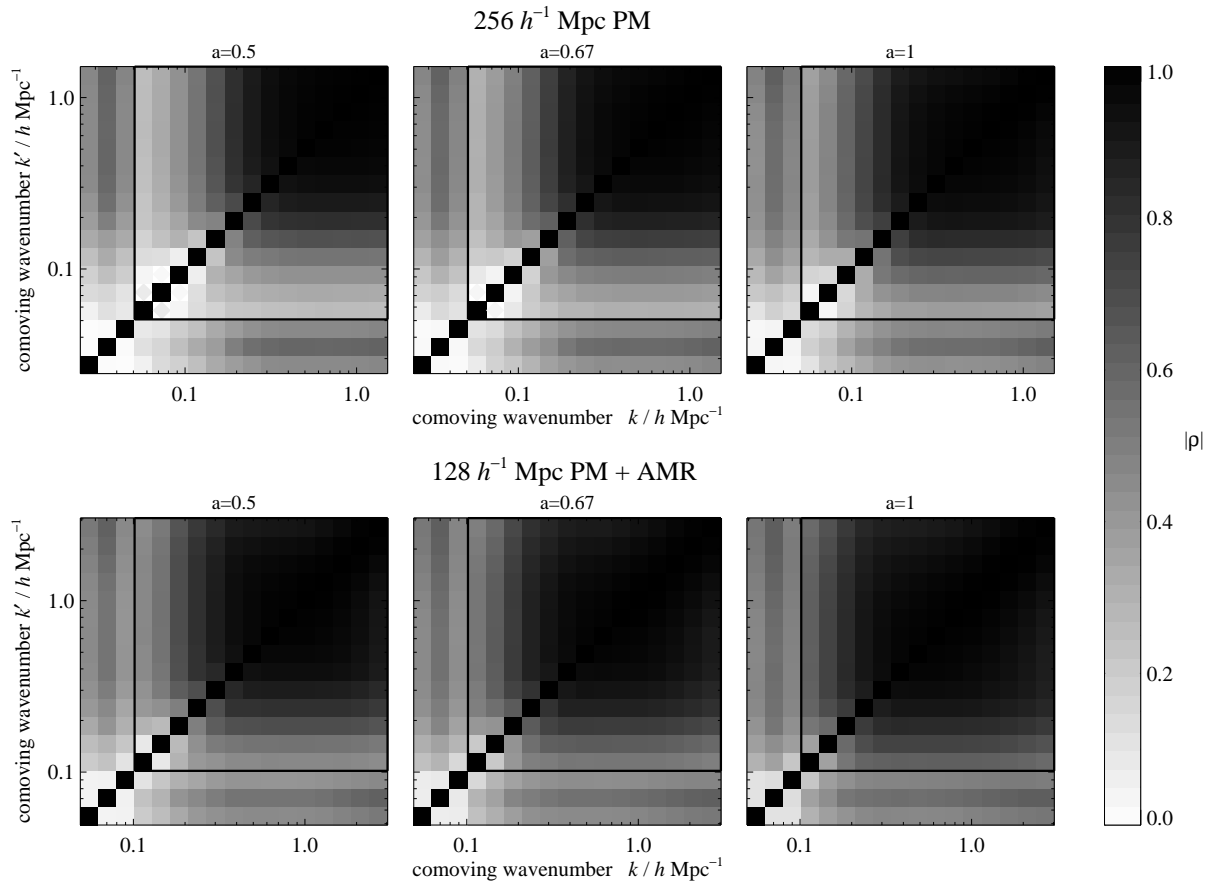}}
\caption{Correlation matrices of estimates of non-linear power using the
  re-weighting scheme of HRS at 3 epochs (left to right $a=0.5$, 0.67 and 1).
  Correlation matrices were estimated separately for each simulation and then
  averaged over simulations.  Results shown are averages over 100 PM
  simulations of each box size and additionally, in the case of the 128~\Mpc\
  boxes, the 25 \art\ simulations.  Greyscale is used to indicate the magnitude
  of the correlations, ranging from 0 (no correlation) to 1 (perfect
  correlation).  A heavy, black border outlines the region of each covariance
  matrix that is unaffected by numerical artefacts from the re-weighting
  scheme.  Bins outside of the bordered area are excluded from further
  analyses.
  \label{fig: covariance}}
\end{figure*}

\subsection{Covariance of power of weighted density}
\label{ssec: weighting}

Following HRS, we define the $i$'th weighted density by
\begin{equation}
  \rho_i(\br) \equiv w_i(\br)\rho(\br).
\end{equation}
We use the minimum variance set of weightings recommended in section~3 of HRS.
In real space the $i$'th weighting has the form
\begin{equation}
  w_i(\br) =
  \sqrt{2}\cos\left[2\upi\left(\bk_i\cdot\br+\frac{1}{16}\right)\right],
  \label{eq: w}
\end{equation}
where
\begin{equation}
  \bk_i = \left\{
    \begin{array}{l@{\quad}l}
      \{1,0,0\} & \mbox{12 weightings} \\
      \{1,1,0\} & \mbox{24 weightings} \\
      \{1,1,1\} & \mbox{16 weightings.} \\
    \end{array}
    \right.     \label{eq: k_i}
\end{equation}
The different weightings are obtained by all possible reflections and rotations
of the components of $\bk$, which yields 26 weightings in total, allowing for
all of the symmetries in equation~(\ref{eq: w}).  A further 26 weightings are
obtained by adding a phase shift of $\upi/2$, i.e.\ $1/16 \to 5/16$ in
equation~(\ref{eq: w}), equivalent to translating one of the coordinates by a
quarter box (because of the symmetry of the weighting functions, only one such
translation yields a distinct weighting).  The estimate of power from this
second set of weightings is predicted by HRS to be highly anti-correlated with
that obtained from the first set of 26 weightings, but they are sufficiently
uncorrelated that including all 52 weightings does yield a better measurement
of the covariance matrix than is obtained from only 26 weightings.  It is
possible to apply more weightings constructed from higher order modes than
those in equation~(\ref{eq: k_i}) but, as argued in section~3.5 of HRS, these
contain progressively less independent information, and yield progressively
less accurate estimates of covariance.

A covariance matrix estimated as the average of $n$ distinct estimates can have
a rank no greater than $n$, a fact also pointed out recently by \citet{PS05}.
The 52 weightings given by equation~(\ref{eq: k_i}) prove sufficient to yield,
for each simulation, a non-singular (no zero eigenvalues), and indeed positive
definite (no negative eigenvalues) estimate of the covariance matrix for the 20
bins of wavenumber used here.

As a practical matter, we implement the re-weighting scheme by weighting
individual particles, before assigning the density to the Fourier mesh.  The
weighted overdensity at point $j$ on the mesh is defined to be
\begin{equation}
  \delta_i(\br_j) \equiv \frac{\rho_i(\br_j)}{\bar{\rho}}-1, \label{eq: deltai}
\end{equation}
where $\bar{\rho}$ is the mean of the \emph{unweighted} density field.  Note
that this is a different convention to that used in HRS, in which the quantity
being transformed is
\begin{equation}
  \Delta\rho_i(\br) \equiv \rho_i(\br) - \bar{\rho}_i(\br),
\end{equation}
where $\bar{\rho}_i(\br)=w_i(\br)$ and $\bar{\rho} \equiv 1$.  The difference
between these two conventions only affects the power on the largest scales
(those wavebands containing modes appearing in equation~\ref{eq: k_i}) and can
most easily (and accurately) be corrected for in Fourier space.  However, since
the covariance on these scales is not correctly reproduced by the re-weighting
method, we simply exclude them from our analyses.

%%%%%%%%%%%%
% FIGURE 3 %
%%%%%%%%%%%%

\begin{figure*}
\centerline{\includegraphics[width=16cm]{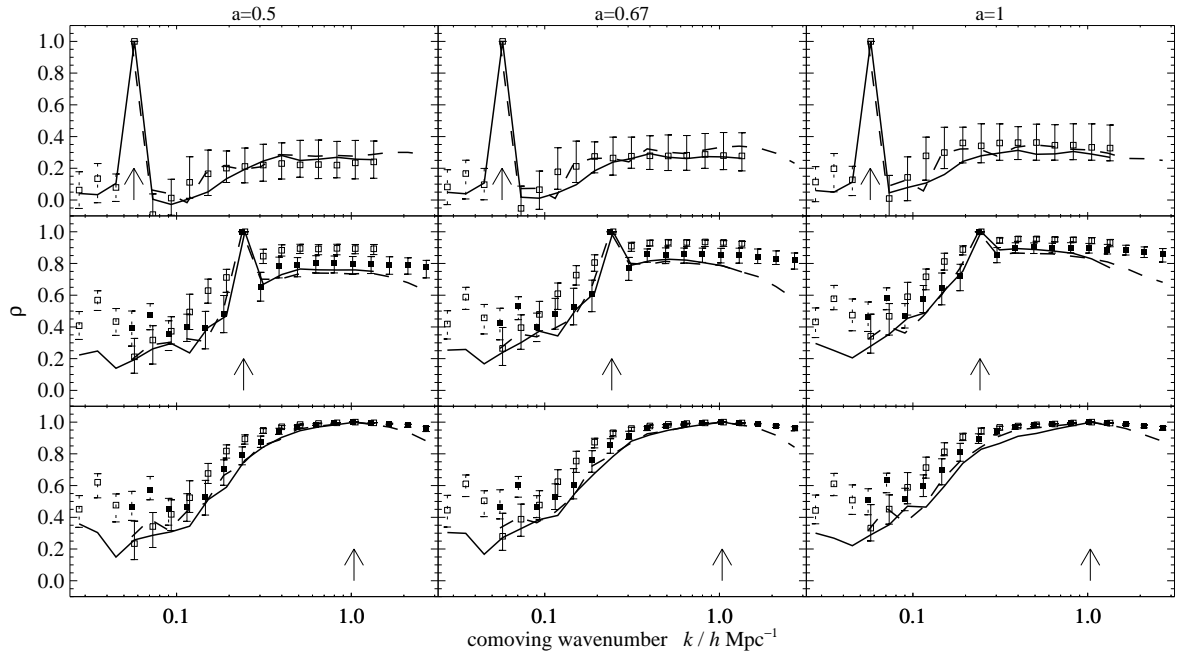}}
\caption{Correlations between estimates of the non-linear power spectrum using
  the re-weighting method, with $\Delta{P_i}$ defined by equation~(\ref{eq:
  strategy 2}).  The three columns show cross-sections through the correlation
  matrices at 3 epochs (left to right $a=0.5$, 0.67 and 1), while each row
  shows a cross-section at a different wavenumber ($k^\prime = 0.057$, 0.242
  and 1.036~\iMpc), marked by the vertical arrow in each panel.  Symbols with
  error bars mark the median and quartiles of the distribution of correlation
  co-efficients measured from each individual simulation for the 256~\Mpc\ PM
  (open symbols) and 128~\Mpc\ PM+\art\ (filled symbols) simulations.  Dotted
  error bars are used for points that are outside of the range of wavenumbers
  for which the covariance is expected to be reliably estimated by the
  re-weighting method.  Solid and dashed lines show the correlations between
  (unweighted) estimates of power using the full ensemble of 256~\Mpc\ PM and
  128~\Mpc\ PM+\art\ simulations, respectively.  Data for the 128~\Mpc\ boxes
  are missing in the top row because this entire cross-section lies outside the
  reliable region for this box size.
  \label{fig: correlation}}
\end{figure*}

%%%%%%%%%%%%
% FIGURE 4 %
%%%%%%%%%%%%

\begin{figure*}
\centerline{\includegraphics[width=16cm]{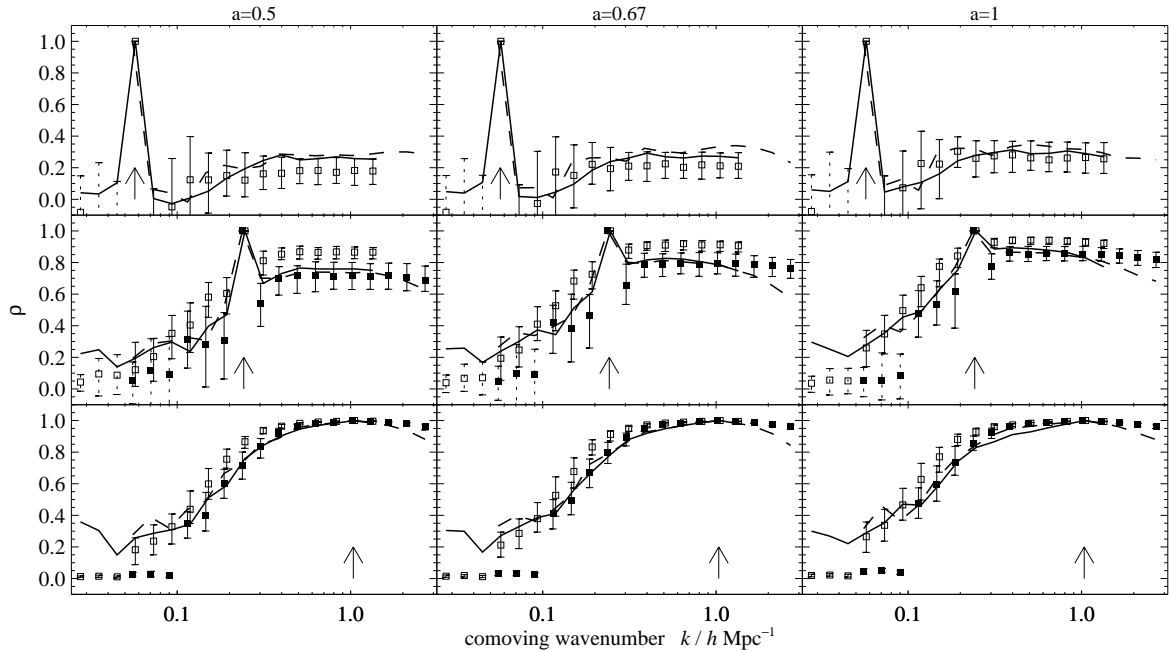}}
\caption{As Fig.~\ref{fig: correlation}, but with $\Delta{P_i}$ defined by
  equation~(\ref{eq: strategy 1}).  The symbols and lines have the same
  meanings as in Fig.~\ref{fig: correlation}.  Together, the two figures
  demonstrate that the different methods give broadly consistent results.
  \label{fig: correlation 2}}
\end{figure*}

%%%%%%%%%%%%
% FIGURE 5 %
%%%%%%%%%%%%

\begin{figure*}
\centerline{\includegraphics[width=16cm]{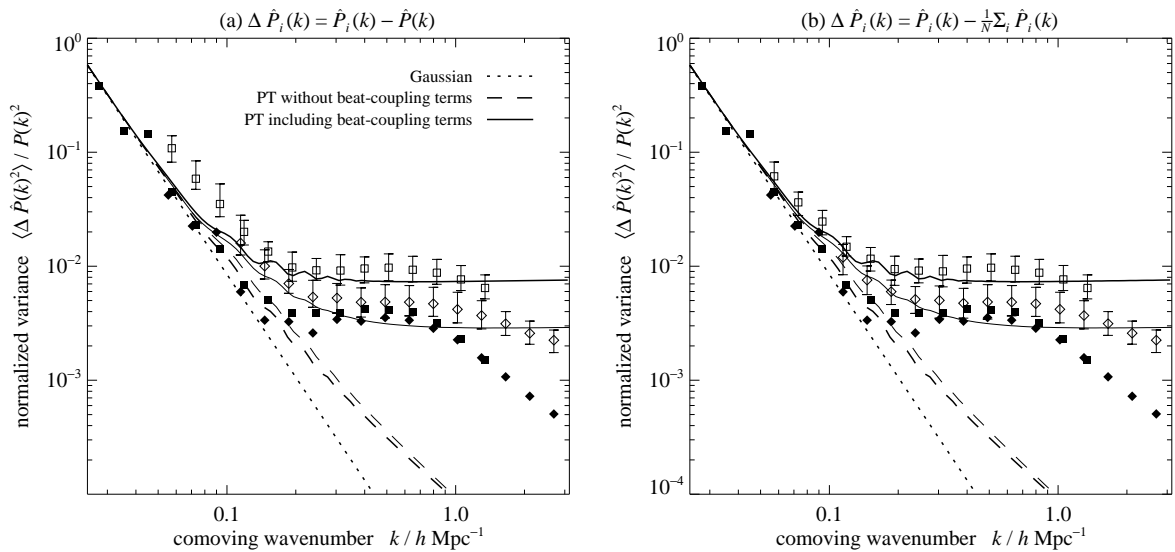}}
\caption{Variance of estimates of power using the re-weighting method with
  $\Delta{P_i}$ defined by (a) equation~(\ref{eq: strategy 2}) and (b)
  equation~(\ref{eq: strategy 1}).  Open symbols show the variance, normalized
  to the square of the power, at $a=1$ in the 256~\Mpc\ simulations (squares)
  and the 128~\Mpc\ simulations (diamonds).  The results for the 128~\Mpc\ box
  size have been shifted vertically by a factor 8 to compensate for the reduced
  density of modes and enable direct comparison between the two box sizes.
  Error bars mark the upper and lower quartiles of the distribution of
  estimates of variance over individual simulations.  Solid symbols give the
  corresponding results measured using the ensemble method (these points are
  the same in both panels).  The solid and dashed curves are predictions from
  perturbation theory, both with (solid) and without (dashed) the beat-coupling
  terms (see HRS for details of the calculation).  The heavier lines are the
  results for the 256~\Mpc\ box size.  The dotted line is the expected variance
  for Gaussian fluctuations.
  \label{fig: variance}}
\end{figure*}

The covariance of power over the ensemble of weighted powers is related to the
true covariance of power by (HRS)
\begin{equation}
  \expect{\Delta{\hat{P}(k)} \Delta{\hat{P}(k^\prime)}} = 2
  \expect{\Delta{\hat{P}_i(k)} \Delta{\hat{P}_i(k^\prime)}}_{i},
%  \expect{\Delta{\hat{P}}_\bk \Delta{\hat{P}}_{\bk^\prime}} = 2
%  \expect{\Delta{\hat{P}}_\bk \Delta{\hat{P}}_{\bk^\prime}}_{i},
\end{equation}
where $\expect{\cdots}_i$ denotes an average over different weightings and the
factor $2$ corrects the covariance of the ensemble to the true covariance of
estimates of power.  The deviation $\Delta\hat{P}_i(k)$ in the power spectrum
of the $i$'th weighted density must be measured relative to some expected or
mean value, and HRS discussed two possibilities.  The first is to measure the
deviation relative to the power spectrum of the unweighted density of the
simulation:
\begin{equation}
  \Delta{\hat{P}_i(k)} \equiv \hat{P}_i(k)-\hat{P}(k).
 \label{eq: strategy 1}
\end{equation}
The second is to measure the deviation relative to the mean of the power
spectra of the weighted densities:
\begin{equation}
  \Delta{\hat{P}_i(k)} \equiv \hat{P}_i(k)-\frac{1}{N}\sum_i{\hat{P}_i(k)}.
 \label{eq: strategy 2}
\end{equation}
The advantage of the first strategy, equation~(\ref{eq: strategy 1}), is that
the power spectrum of the unweighted density is, by symmetry, the most accurate
estimate of power in a simulation, so the statistical uncertainty is
potentially least in this case.  However, the power spectra of weighted
densities are likely to be slightly biased relative to the power spectrum of
the unweighted density, because weighting the density effectively smooths the
power, which biases it if power is other than a linear function of wavenumber.
The advantage of the second strategy, equation~(\ref{eq: strategy 2}), is that
it removes this slight bias, so the systematic uncertainty is potentially
smaller in this case.

The lower panel of Fig.~\ref{fig: power} shows the median and quartiles of the
distribution of the deviations between the averaged power spectra of weighted
densities and the power spectrum of the unweighted density of each simulation
(c.f.\ equation~20 of HRS).  The two agree well on small scales, but on larger
scales they can differ by up to 20 percent in extreme cases.

In the following sections, we show results for both strategies,
equations~(\ref{eq: strategy 1}) and (\ref{eq: strategy 2}), and find that the
two strategies give consistent results.  However, we do find that
equation~(\ref{eq: strategy 1}) gives variances that are slightly but
systematically higher than those from equation~(\ref{eq: strategy 2}), which we
attribute to the systematic bias between the power spectra of weighted
densities versus the power spectrum of the unweighted density.  For the
purposes of computing information, Section~\ref{sec: results}, we therefore
choose the second strategy, equation~(\ref{eq: strategy 2}).

Each set of weightings in equation~(\ref{eq: k_i}) is generated from a
different wavenumber $k_i=|\bk_i|$, so we expect estimates of power from each
set to be biased in a slightly different way.  This bias is small; nevertheless
it is preferable to estimate the covariance matrix separately from each set of
weightings and then combine them to obtain a single estimate of the covariance
matrix, weighting by the number of weightings in each set.  This is the
procedure that we adopt in Section~\ref{sec: results}.

\subsection{Tests}
\label{ssec: tests}

In this section we describe several tests of the measurement of covariance
of power.  In Section~\ref{sssec: correlations} we show that the weightings and
ensemble methods give consistent results for the correlation coefficients of
band-powers.  By contrast, in Section~\ref{sssec: variance} we show that the
two methods give substantially different results for the variance of power at
non-linear scales.  In Section~\ref{sssec: gaussian} we check the assumption
that the distribution of estimates of power is (thanks to the central limit
theorem) adequately Gaussian.

\subsubsection{Correlations in the power spectrum}
\label{sssec: correlations}

Fig.~\ref{fig: covariance} shows the matrix of correlation coefficients,
\begin{equation}
  \rho_{kk^\prime} \equiv \frac{\expect{\Delta{P_k}\Delta{P_{k^\prime}}}}
      {\sqrt{\expect{\Delta{P_k}^2}\expect{\Delta{P_{k^\prime}}^2}}},
%  \rho(k,k^\prime) \equiv \frac{\expect{\Delta{P(k)}\Delta{P(k^\prime)}}}
%  {[\expect{\Delta{P(k)}^2}\expect{\Delta{P(k^\prime)}^2}]^{1/2}},
\end{equation}
of estimates of the non-linear power spectrum at three epochs, measured using
the re-weighting scheme outlined in Section~\ref{ssec: weighting}.  Each matrix
is the average result from 100 individual PM simulations and, in the case of
the 128~\Mpc\ boxes, 25 \art\ simulations.  The final epoch ($a=1$) can be
directly compared to fig.~2 of Paper~I, in which we show the results from the
ensemble method.  We expect numerical artefacts from the re-weighting to be
restricted to wavenumbers $k \leq \sqrt{3}k_{\rm b}$, the highest wavenumber
used in the weighting functions.  Fig.~\ref{fig: covariance} shows that this is
indeed the case: the degree of correlation changes abruptly between bins
containing wavenumbers inside and outside this limit.  In the following
analysis, we use only bins with wavenumbers a factor of 2 away from the
fundamental mode of the box, $k \geq 2 k_{\rm b}$.

Fig.~\ref{fig: correlation} shows three cross-sections through each of the
correlation matrices in Fig.~\ref{fig: covariance}.  The data plotted are the
medians and quartiles of the distribution over the independent realizations.
The measurements from the re-weighting and ensemble methods are generally
consistent, although the re-weightings method tends to yield somewhat higher
correlations at smaller scales, and higher for the $256 \, h^{-1} \mbox{Mpc}$
boxes than the $128 \, h^{-1} \mbox{Mpc}$ boxes.  The scatter between
individual simulations is considerable, especially where the correlation
coefficient is small.

Fig.~\ref{fig: correlation 2} shows the same results, but with the deviations
$\Delta\hat{P}_i(k)$ in the power spectra of weighted densities being measured
relative to the power spectrum of the unweighted density, equation~(\ref{eq:
strategy 1}), as opposed to the mean of the power spectra of weighted
densities, equation~(\ref{eq: strategy 2}).  The correlation coefficients are
similar to those shown in Figure~\ref{fig: correlation}, as they should be.

\subsubsection{Variance of the power spectrum}
\label{sssec: variance}

While the re-weighting and ensemble methods yield consistent results for the
correlation matrix of non-linear power, the variance of (and hence the
information contained in) the non-linear power spectrum is an altogether
different matter.

Fig.~\ref{fig: variance} shows the variance, normalized to the square of the
unweighted power spectrum, estimated using the re-weighting method.  We show
results for both equation~(\ref{eq: strategy 1}) and equation~(\ref{eq:
strategy 2}).  Notice that, as expected, equation~(\ref{eq: strategy 1})
overestimates the variance of power on large scales.  On small scales, however,
the results are entirely consistent.

For Gaussian fluctuations, the normalized variance equals $2/N_k$
(equation~\ref{eq: variance}), where $N_k$ is the number of modes in a given
wavenumber bin.  For a periodic box, the number of modes in a set of
logarithmically-spaced bins increases with central wavenumber as $N_k \propto
k^3$.  The 128~\Mpc\ simulations have fewer modes (by a factor of 8) at a given
wavenumber, so the results for this box size have been shifted vertically down
by this factor to allow a direct comparison between the results for the two box
sizes.  As in Fig.~\ref{fig: correlation}, the data shown in Fig.~\ref{fig:
variance} are medians over many individual simulations, with error bars marking
the quartiles of the distribution.  The results for the 128~\Mpc\ PM
simulations and the 128~\Mpc\ \art\ simulations are consistent, so we combine
them into a single set of points for clarity (a figure showing the \art\
results alone appears as fig.~2 of HRS).

The two different box sizes yield consistent results when the ensemble method
is used.  For the re-weighting method, on the other hand, there are clear
discrepancies, both between the re-weighting method and the ensemble method and
between the two box sizes.  At translinear and non-linear scales, the variance
measured by re-weighting is significantly higher than that measured for the
ensemble, particularly for the larger box size, and the discrepancy grows ever
larger at smaller scales.

Fig.~\ref{fig: variance} also shows the predictions of perturbation theory.
The calculation of these curves is described in section~4.3 of HRS.
Perturbation theory helps to explain the discrepancies both between the results
of the ensemble and re-weighting methods, and between the two different box
sizes.  The variance measured by the re-weighting technique departs from the
ensemble result where the (constant) beat-coupling term (equation~94 of HRS)
becomes the dominant source of variance.  The source of this term -- coupling
of closely-spaced Fourier modes to the large-scale beat mode between them -- is
discussed in section~4 of HRS.  The magnitude of the beat-coupling term is
proportional to the amplitude of the power spectrum on roughly the size of the
box, which explains why different sizes of simulation box yield systematically
different estimates of the small-scale variance.  Perturbation theory correctly
predicts the magnitude of the discrepancy between the small-scale variance
estimated from the two different sizes of simulation.

Note that perturbation theory fails to reproduce the correct non-linear
variance for the ensemble method (dashed lines), presumably because
perturbation theory breaks down in the highly non-linear regime.  This may be
responsible for the small discrepancies between the results of the
re-weightings method and the theoretical curves in Fig.~\ref{fig: variance}.

%%%%%%%%%%%%
% FIGURE 6 %
%%%%%%%%%%%%

\begin{figure}
\centerline{\includegraphics[width=8cm]{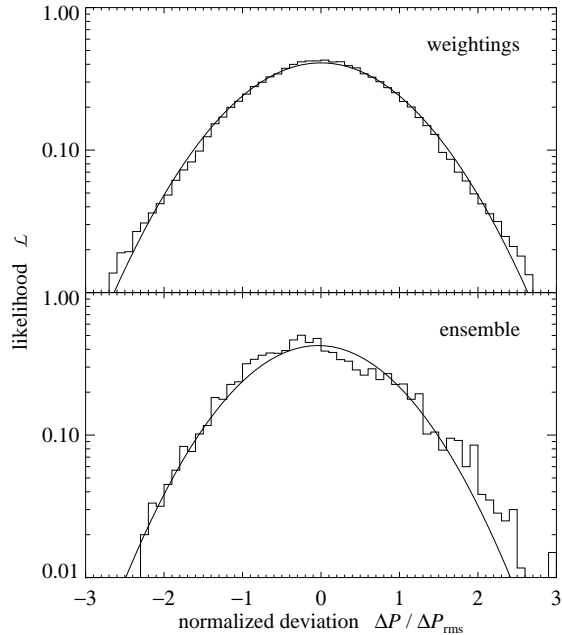}}
\caption{Distributions of estimates of non-linear power, equation~(\ref{eq:
  strategy 2}), using the re-weighting method (top panel) and the ensemble
  method (bottom panel).  Individual histograms for all bands with wavenumbers
  $k > 0.2$~\Mpc, containing at least 2000 Fourier modes, are scaled to unit
  variance and stacked.  The smooth curves are Gaussian fits to the data,
  assuming Poisson weighting of the counts in each bin.
  \label{fig: gaussian}}
\end{figure}

%%%%%%%%%%%%
% FIGURE 7 %
%%%%%%%%%%%%

\begin{figure*}
\centerline{\includegraphics[width=16cm]{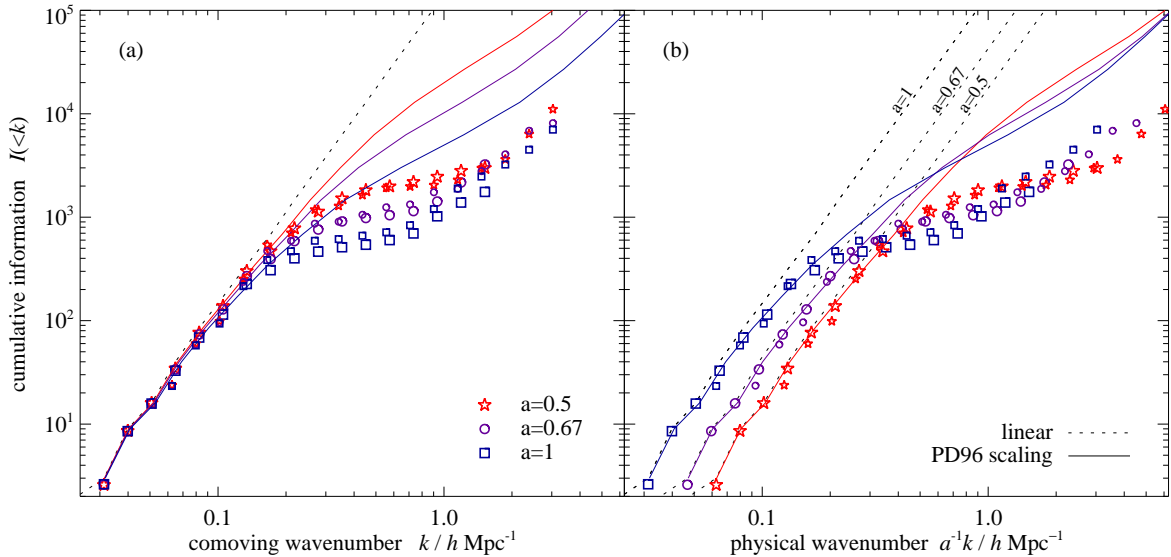}}
\caption{Cumulative information in the non-linear power spectrum at 3 epochs
  (top to bottom: $a=0.5$, 0.67 and 1), as a function of (a) comoving and (b)
  physical wavenumber.  Large symbols are points derived from the 256~\Mpc\ PM
  simulations and small simulations are for the 128~\Mpc\ PM+\art\ boxes.  The
  results for the 128\Mpc\ boxes have been shifted vertically by a factor 8 to
  account for the higher density of modes at a given comoving $k$ and allow
  direct comparison with the larger box size.  The dotted line marks the
  expected amount of information for Gaussian fluctuations and the solid curves
  are the result of applying the PD96 scaling to the dotted line at each epoch.
  This is a revised version of fig.~3 in Paper~I (see text for details).
  \label{fig: info}}
\end{figure*}

\subsubsection{Distribution of estimates of power}
\label{sssec: gaussian}

In Section~\ref{sec: info} we asserted that, thanks to the central limit
theorem, estimates of power should be Gaussianly distributed about their
expectation value, even in the highly non-linear regime.  Since our analysis
relies on the validity of this assumption it is worth putting to the test.

Fig.~\ref{fig: gaussian} shows the distribution of deviations of estimates of
power using both the re-weighting method and the ensemble method for
wavenumbers in the non-linear regime (which we take to be $k > 0.2$~\Mpc).
Individual histograms for each waveband have been scaled to unit variance and
stacked to produce a single distribution for each method.  The estimates of
power from the re-weighted simulations are indeed distributed close to
Gaussianly.  Assuming Poisson statistics, the value of $\chi^2$ for the fit is
5.84 per degree of freedom, which seems reasonable, given the high degree of
correlation between estimates of non-linear power.  For the ensemble method,
the distribution is also close to Gaussian, although there are deviations from
Gaussianity -- in particular the presence of a sharper than Gaussian peak and a
tail of large, positive deviations that cause the actual mean and variance of
power to be slightly larger than the fitted values.  The value of $\chi^2$ in
this case is 3.86 per degree of freedom, for Poisson statistics.

%%%%%%%%%%%%%%%%%%%%%%%%%%%%%%%%%%%%%%%%%%%%%%%%%%%%%%%%%%%%%%%%%%%%%%%%%%%%%%%
%%%%%%%%%%%%%%%%%%%%%%%%%%%%%%%%%%%%%%%%%%%%%%%%%%%%%%%%%%%%%%%%%%%%%%%%%%%%%%%

\section{Information content of the non-\\ linear power spectrum}
\label{sec: results}

In the previous section, we showed that measuring the covariance of non-linear
power by re-weighting an individual simulation yields substantially different
results at non-linear scales than is found from the scatter over an ensemble of
simulations.  The discrepancy is consistent with the explanation proposed by
HRS: beat-coupling between the covariance on non-linear scales and the power on
large scales.  The difference in covariance between the two methods translates
directly into a difference in the quantity of information in the non-linear
power spectrum.

\subsection{Method}
\label{ssec: method}

We decorrelate estimates of the (log-) amplitude for each simulation
individually, using the covariance matrix estimated using the re-weighting
method.  As our fiducial power spectrum $P_k$ in equation~(\ref{eq: band
powers}) we use the true (unweighted) power in each simulation.  The
re-weighting method restricts the range of wavenumbers for which the covariance
matrix -- and hence the Fisher information -- can be reliably measured to $k >
\sqrt{3}k_{\rm b}$.  Since our purpose is to measure the \emph{cumulative}
information, equation~(\ref{eq: IPk}), up to some wavenumber $k_{\rm max}$, the
contribution from wavenumbers smaller than this limit must be taken into
account.  For the 256~\Mpc\ boxes we assume that the excluded bins contain the
expected amount of information for Gaussian fluctuations ($N_k/2$, where $N_k$
is the number of Fourier modes in those bins).  That this is a reasonable
assumption is confirmed by the fact that the first few bins for which we do
have measurements of the variance using the re-weighting method follow the
Gaussian expectation closely (see the top panel of Fig.~\ref{fig: variance}).
For the 128~\Mpc\ boxes there is a further complication.  The lower limit on
the wavenumbers accessible using the re-weighting method brings us into the
regime in which non-linear effects start to become important.  The most
reasonable thing to do here would seem to be to use the results from the
256~\Mpc\ boxes to estimate the quantity of missing information.  Although
there are clearly systematic differences between the results of the
re-weighting method for different box sizes, these are small at the scales in
question.

%%%%%%%%%%%%
% FIGURE 8 %
%%%%%%%%%%%%

\begin{figure*}
\centerline{\includegraphics[width=16cm]{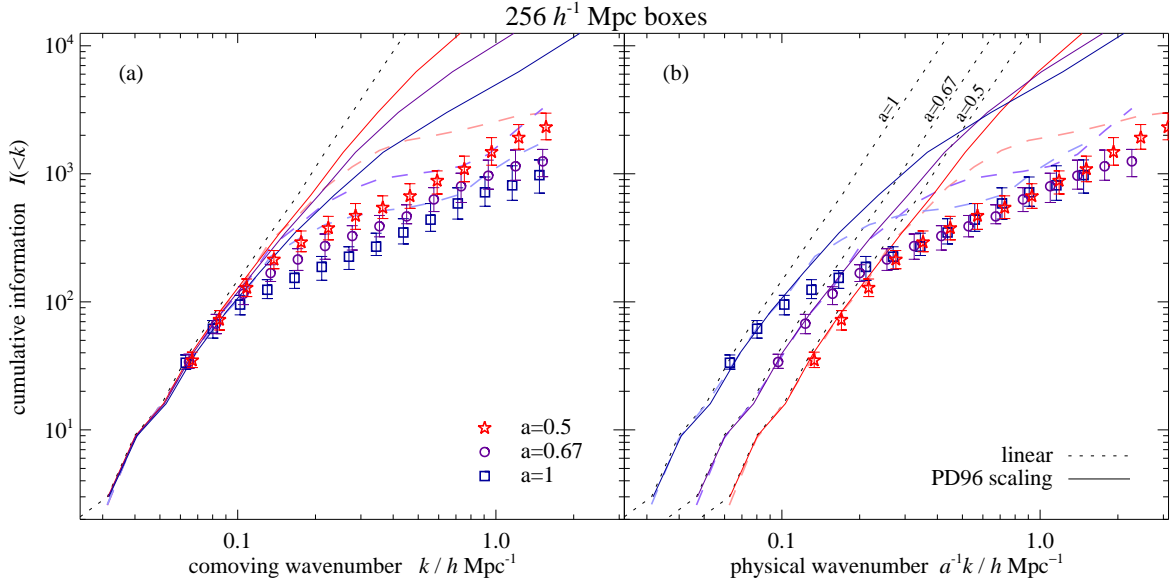}}
\caption{Cumulative information, as a function of (a) comoving and (b) physical
  wavenumber, for the same 3 epochs as Fig.~\ref{fig: info}.  Points with error
  bars are medians and quartiles of the results measured using the re-weighting
  method on 100 PM simulations with a box size of 256~\Mpc.  For clarity, the
  three sets of points have been artificially separated by a small factor in
  wavenumber, with the $a=0.67$ points having the correct wavenumber.  For
  comparison, the ensemble results from Fig.~\ref{fig: info} are shown as
  light, dashed lines.  The dotted line marks the expected amount of
  information for Gaussian fluctuations and the solid curves are the result of
  applying the PD96 scaling to the dotted line at each epoch.
  \label{fig: info 256}}
\end{figure*}

%%%%%%%%%%%%
% FIGURE 9 %
%%%%%%%%%%%%

\begin{figure*}
\centerline{\includegraphics[width=16cm]{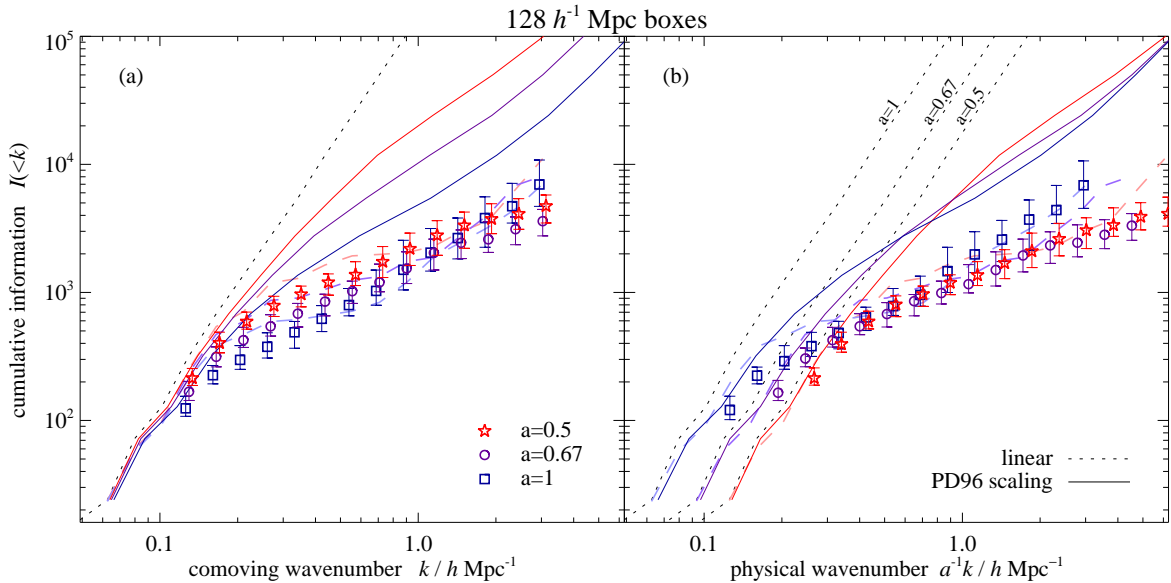}}
\caption{Cumulative information, as Fig.~\ref{fig: info 256}, for 100 PM
  simulations and 25 \art\ simulations with a box size of 128~\Mpc.  The
  results have been shifted vertically by a factor of 8 to enable direct
  comparison with the larger box size.  The symbols and lines have the same
  meanings as in Fig.~\ref{fig: info 256}, but note the different axis ranges.
  \label{fig: info 128}}
\end{figure*}

\subsection{Results}
\label{ssec: results}

\subsubsection{Ensemble method}
\label{sssec: results 1}

Fig.~\ref{fig: info} shows the cumulative information as a function of
wavenumber at 3 epochs ($a=0.5$, 0.67 and 1) for the ensemble method of
estimating the covariance of power.

The curves in Fig.~\ref{fig: info} differ somewhat from those in fig.~3 of
Paper~I.  In the previous paper, we (incorrectly) decorrelated the power
spectrum \emph{prior} to multiplying by the partial derivatives in
equation~(\ref{eq: IPk}).  As discussed in Section~\ref{ssec: decorrelating},
this is only a good approximation to the exact result provided that the
band-power windows are narrow in $k$, which is a poor assumption in our case
because of the existence of broad correlations in the power spectrum.  The
curves in Fig.~\ref{fig: info} (and Figs.~\ref{fig: info 256} and \ref{fig:
info 128} later) were produced using the exact formalism set out in
Section~\ref{ssec: decorrelating} of the present paper.  We still find that the
ensemble method yields very little independent information in the translinear
regime ($k \simeq 0.2$--0.8~\Mpc), over and above that in the linear regime,
although the flatness is not as pronounced as suggested by our previous
results.  In fact, the cumulative information increases by a factor of
approximately 2 over the aforementioned range of wavenumbers (at $a=1$), still
much less than the factor of $k^3 \simeq 64$ expected for linear fluctuations.
There remains a sudden upturn in the cumulative information on small scales,
implying that the power spectrum at fully non-linear scales contains unique
information about the amplitude of the initial power spectrum, that is not
found in the present day linear power spectrum.

In Paper~I, we interpreted the decrease in information in the translinear
regime as the result of rapid transfer of information from larger to smaller
scales.  An alternative explanation, which was mentioned in Paper~I but
discarded as being contrived, is that information is temporarily diverted into
higher order statistics, such as the bispectrum, in the translinear regime.
Since filamentary structures are more readily described by higher-order
statistics than by the power spectrum alone, it is entirely plausible that, on
translinear scales, the bispectrum does contain information not present in the
power spectrum, about the initial conditions of structure formation, and that
this information is somehow returned to the power spectrum on smaller scales.

We argued in Paper~I that, if information is conserved overall then, under the
assumption of stable clustering, the amount of information up to a given
physical (as opposed to comoving) wavenumber ought to be independent of time in
the fully non-linear regime.  The right panel of Fig.~\ref{fig: info} shows the
cumulative information for the same three epochs, plotted as a function of
physical wavenumber $k/a$.  The results are consistent with the hypothesis that
information is largely conserved by non-linear evolution.

\subsubsection{Re-weighting method}
\label{sssec: results 2}

In Figs.~\ref{fig: info 256} and \ref{fig: info 128} we compare the results
from the ensemble method with those from the re-weighting method.  For the
re-weighting method we show the median and quartiles of the distribution of
results from the individual simulations.  Qualitatively, the behaviour of the
cumulative information -- as a function of both wavenumber and cosmic epoch --
for the re-weighting method is similar to that for the ensemble method.
However, the flattening in the translinear regime is less pronounced than for
the ensemble case and there is no clear evidence for an upturn on small scales,
although the curves for the re-weighting method follow the average slope of
those from the ensemble method.  Overall, the information measured using the
re-weighting method is considerably less than when the ensemble method is used.
Beat-coupling to large scales prevents much of the information that is, in
principle, contained in the power spectrum from being extracted when the
re-weighting method is used.  It is also worth noting that the magnitude of the
beat-coupling effect depends on the size of the simulation, so that the two
different box sizes are no longer in complete agreement on small scales.  The
128~\Mpc\ boxes (Fig.~\ref{fig: info 128}) are in closer agreement with the
ensemble method, as expected.

As predicted by perturbation theory, the beat-coupling contribution to the
covariance, which is a factor $\sim P(2k_{\rm b})/P(k)$ larger than the other
terms at a given $k$, becomes increasingly dominant at smaller scales.  For the
power spectrum considered here, the contribution from beat-coupling also
increases with cosmic epoch.  We would therefore not expect the cumulative
information, measured using the re-weighting method, up to a given physical
wavenumber to be necessarily constant over time, even in the stable clustering
regime.

%%%%%%%%%%%%%%%%%%%%%%%%%%%%%%%%%%%%%%%%%%%%%%%%%%%%%%%%%%%%%%%%%%%%%%%%%%%%%%%
%%%%%%%%%%%%%%%%%%%%%%%%%%%%%%%%%%%%%%%%%%%%%%%%%%%%%%%%%%%%%%%%%%%%%%%%%%%%%%%

\section{Summary}
\label{sec: summary}

This paper extends and expands on the results reported in Paper~I concerning
the Fisher information contained in the non-linear power spectrum about the
amplitude of the initial (post-recombination) linear power.

In Paper~I, we measured the covariance of power from the scatter in power over
a large ensemble of simulations.  Here we reported measurements of covariance
of power from both the ensemble method and a new method, described in a
companion paper (HRS), in which smoothly varying weighting functions are
applied to each simulation to yield a separate estimate of the covariance of
power for each simulation.

We have shown that the two methods yield substantially different estimates for
the covariance of power at non-linear scales.  This does not mean, however,
that one or other of the methods is incorrect.  Rather, it turns out that
measuring the covariance of power is a more subtle problem than we had
previously suspected.  Beat-coupling -- the process whereby the covariance
between Fourier modes separated by a small wavevector couple by gravitational
growth to the large-scale beat mode between them -- dominates the covariance on
non-linear scales.  We compared our results to a calculation using perturbation
theory (HRS) and found that the theory explains qualitatively the features of
beat-coupling, seen in the simulations.

Beat-coupling contributions to the covariance of power occur whenever Fourier
modes have a finite width, as opposed to being delta-functions at discrete
wavevectors.  As argued by HRS, this means that beat-coupling is likely to be
important in real galaxy surveys.  The ensemble method, on the other hand,
measures covariances between the amplitudes of Fourier modes with precisely
defined (delta function) wavenumbers.  If information is conserved by
non-linear evolution then it is this quantity that, overall, remains invariant
with time.

Theory predicts that the effects of beat-coupling will be greatest when the
largest scales in a survey are close to the peak of the power spectrum ($k
\simeq 0.016$~\iMpc; $\upi/k \sim 200$~\Mpc).  If this is true then our results
suggest the covariance of power in such a survey will be dominated by
beat-coupling on small scales and, counter-intuitively, will be sensitive to
the power on the largest scales in the survey, leading to a reduction in the
amount of information extractable from the power spectrum.  As pointed out in
HRS, the best way to test this is using mock galaxy catalogues drawn from a
single large simulation using the same selection function as the survey
observations.

%%%%%%%%%%%%%%%%%%%%%%%%%%%%%%%%%%%%%%%%%%%%%%%%%%%%%%%%%%%%%%%%%%%%%%%%%%%%%%%
%%%%%%%%%%%%%%%%%%%%%%%%%%%%%%%%%%%%%%%%%%%%%%%%%%%%%%%%%%%%%%%%%%%%%%%%%%%%%%%

\section*{Acknowledgments}

We are grateful to Anatoly Klypin and Andrey Kravtsov for making the MPI
implementation of \art\ available to us and for help with its application.  We
also thank Nick Gnedin, Mathias Zaldarriaga, Rom\'an Scoccimarro and Max
Tegmark for useful discussions and the anonymous referee for his/her comments.
This work was supported by NSF grant AST-0205981 and by NASA ATP award
NAG5-10763.  \grafic\ is part of the {\sc cosmics} package, which was developed
by Edmund Bertschinger under NSF grant AST-9318185.  Some of the simulations
used in this work were performed at the San Diego Supercomputer Center using
resources provided by the National Partnership for Advanced Computational
Infrastructure under NSF cooperative agreement ACI-9619020.

\bibliographystyle{my_mn2e}
\bibliography{info}

\appendix

\bsp

\label{lastpage}

\end{document}